# A Note on Embedding of M-Theory Corrections into Eleven-Dimensional Superspace[1]

Hitoshi Nishino[2]

*Department of Physics*
*University of Maryland*
*College Park, MD 20742-4111*

and

Subhash Rajpoot[3]

*Department of Physics & Astronomy*
*California State University*
*Long Beach, CA 90840*

## Abstract

By analyzing eleven-dimensional superspace fourth-rank superfield strength $F$-Bianchi identities, we show that M-theory corrections to eleven-dimensional supergravity can not be embedded into the mass dimension zero constraints, such as the $(\gamma^{ab})_{\alpha\beta} X_{ab}{}^c$ or $i(\gamma^{a_1\cdots a_5})_{\alpha\beta} X_{a_1\cdots a_5}{}^c$ -terms in the supertorsion constraint $T_{\alpha\beta}{}^c$. The only possible modification of superspace constraint at dimension zero is found to be the scaling of $F_{\alpha\beta cd}$ like $F_{\alpha\beta cd} = (1/2)(\gamma_{cd})_{\alpha\beta} e^\Phi$ for some real scalar superfield $\Phi$, which alone is further shown not enough to embed general M-theory corrections. This conclusion is based on the dimension zero $F$-Bianchi identity under the two assumptions: (i) There are no negative dimensional constraints on the $F$-superfield strength: $F_{\alpha\beta\gamma\delta} = F_{\alpha\beta\gamma d} = 0$; (ii) The supertorsion $T$-Bianchi identities and $F$-Bianchi identities are not modified by Chern-Simons terms. Our result can serve as a powerful tool for future exploration of M-theory corrections embedded into eleven-dimensional superspace supergravity.

---

[1]This work is supported in part by NSF grant # PHY-98-02551.
[2]E-Mail: nishino@nscpmail.physics.umd.edu
[3]E-Mail: rajpoot@csulb.edu

## 1. Introduction

If M-theory [1] unifies superstring theories [2], such as type-I, heterotic and type-IIA superstring theories, and its low energy limit is described by eleven-dimensional (11D) supergravity [3], it is natural to expect that there must be high energy corrections to 11D supergravity compatible with supergravity formulation itself, just as superstring corrections can be embedded into 10D supergravity [2][4]. Based on this principle, there have been attempts to embed such M-theory corrections into 11D supergravity, such as in component formulation [5], as well as in superspace formulation [6][7][8][9].

In superspace formulation [7][8], it is so far commonly believed or expected that such M-theory corrections are most likely embedded into the generalized symmetric matrix components of the mass dimension zero ($d = 0$) supertorsion component $T_{\alpha\beta}{}^c$, such as $(\gamma^{ab})_{\alpha\beta} X_{ab}{}^c$ or $i(\gamma^{a_1\cdots a_5})_{\alpha\beta} X_{a_1\cdots a_5}{}^c$ with some appropriate superfields $X_{ab}{}^c$ or $X_{a_1\cdots a_5}{}^c$ [7][8][10]. However, it is not clear whether we need only the corrections of $T_{\alpha\beta}{}^c$ or $F_{\alpha\beta cd}$ at $d = 0$ alone for embedding M-theory corrections, or we also need any negative dimensional ($d < 0$) constraints, such as $F_{\alpha\beta\gamma d}$ or $F_{\alpha\beta\gamma\delta}$ for such modifications for $F$-Bianchi identities (BIs).

In this Letter, we will present a 'no-go theorem' for embedding M-theory corrections into constraints for superspace BIs with modified constraints only at $d = 0$. We will show that the $d = 0$ corrections of constraints are not enough for embedding M-theory corrections into 11D superspace supergravity. Our conclusion is based on two assumptions: (i) All the $F$-superfield strength constraints at $d < 0$ vanish; (ii) The supertorsion $T$-BIs and $F$-BIs are not modified by Chern-Simons terms. In addition to these assumptions, our conclusion also relies on the so-called 'conventional constraints' that relate various superfields in the most general expansions of the superspace derivatives: $E_\alpha$ and $E_a$ [9]. These conventional constraints are restrictive, e.g., the one-gamma term in $T_{\alpha\beta}{}^c$ is only the standard one: $i\, (\gamma^c)_{\alpha\beta}$, while the two-gamma term $(\gamma^{de})_{\alpha\beta} X_{de}{}^c$ and the five-gamma term $(\gamma^{d_1\cdots d_5})_{\alpha\beta} X_{d_1\cdots d_5}{}^c$ corrections are of a general form. Furthermore, the $X$'s themselves are restricted e.g., $X_{ab}{}^b = 0$, etc., as will be shown later.

A statement for the necessity of the $F$-constraints at $d < 0$ has been given in [8], but without any proof. In the present paper, we provide explicit evidence for that claim. By studying the $F$-BI at $d = 0$, we show that, as long as the $F$-constraints at $d < 0$ are absent, there can be *no* such corrections as $(\gamma^{de})_{\alpha\beta} X_{de}{}^c$ or $(\gamma^{d_1\cdots d_5})_{\alpha\beta} X_{d_1\cdots d_5}{}^c$ possible in $T_{\alpha\beta}{}^c$ other than the standard one-gamma term that can embed M-theory corrections.

## 2. Solving $F$-BI at $d = 0$

We first give the most important part of our results here, namely we analyze whether the $d = 0$ $F$-BI of the type $(\alpha\beta\gamma\delta e)$:

$$\tfrac{1}{6}\nabla_{(\alpha} F_{\beta\gamma\delta)e} + \nabla_e F_{\alpha\beta\gamma\delta}$$
$$- \tfrac{1}{6} T_{e(\alpha|}{}^f F_{f|\beta\gamma\delta)} - \tfrac{1}{6} T_{e(\alpha|}{}^\eta F_{\eta|\beta\gamma\delta)} - \tfrac{1}{4} T_{(\alpha\beta|}{}^f F_{f|\gamma\delta)e} - \tfrac{1}{4} T_{(\alpha\beta|}{}^\eta F_{\eta|\gamma\delta)e} \equiv 0 \qquad (2.1)$$

allows any non-trivial solution possible for embedding M-theory corrections [8]. For simplicity (as the assumption of our 'no-go theorem'), we put the $d < 0$ $F$-constraints to be zero:

$$F_{\alpha\beta\gamma\delta} = 0 \ , \qquad F_{\alpha\beta\gamma d} = 0 \ . \qquad (2.2)$$

Accordingly, there remains only one term in (2.1) at $d=0$:

$$T_{(\alpha\beta|}{}^f F_{f|\gamma\delta)e} \equiv 0 \ . \tag{2.3}$$

It is now a purely algebraic question whether there can be any non-trivial solution to (2.3), when we postulate

$$T_{\alpha\beta}{}^c = i(\gamma^c)_{\alpha\beta} + \tfrac{1}{2}(\gamma^{ab})_{\alpha\beta} X_{ab}{}^c + \tfrac{i}{120}(\gamma^{a_1\cdots a_5})_{\alpha\beta} X_{a_1\cdots a_5}{}^c \ , \tag{2.4}$$

for the $d=0$ supertorsion constraint for $T_{\alpha\beta}{}^c$ [8]. Here we have no corrections for the first one-gamma term, due to the 'conventional constraints' relating $E_\alpha$ and $E_a$ as in eqs. (24) and (25) in [9]. On the other hand, $X_{ab}{}^c$ and $X_{a_1\cdots a_5}{}^c$ are some appropriate superfields that can possibly embed M-theory corrections [7][8]. Accordingly, we need to put the most general corrections also into the $d=0$ $F$-constraint[4]

$$F_{\alpha\beta cd} = i(\gamma^e)_{\alpha\beta} U_{ecd} + \tfrac{1}{2}(\gamma^{ef})_{\alpha\beta} U_{efcd} + \tfrac{i}{120}(\gamma^{[5]})_{\alpha\beta} U_{[5]cd} \ . \tag{2.5}$$

In particular, the lowest order on-shell physical superfield constraint in [11] corresponds to the special case of

$$F_{\alpha\beta cd} = +\tfrac{1}{2}(\gamma_{cd})_{\alpha\beta} \ , \qquad U_{abcd} = +\tfrac{1}{4}\eta_{a[c}\eta_{d]b} \ , \tag{2.6}$$

with $U_{abc} = U_{a_1\cdots a_5 bc} = 0$. Since we are considering M-theory corrections, these $X$ and $U$-superfields can be dealt as perturbation, namely we can first consider the satisfaction of the BI (2.3) at the linear order, temporarily ignoring the bilinear order (cross terms). For this reason, we concentrate on the analysis at the linear order in terms of $X$'s and $U$'s.

Based on the fluctuation analysis [9], we can impose on the $X$'s the following 'conventional constraints'

$$X_{ab}{}^b = 0 \ , \qquad X_{[abc]} = 0 \ , \qquad X_{a_1\cdots a_4 b}{}^b = 0 \ , \qquad X_{[a_1\cdots a_5 b]} = 0 \ . \tag{2.7}$$

Under these constraints, $X_{ab}{}^c$ has only 429 degrees of freedom, while $X_{[5]}{}^c$ has 4,290 degrees of freedom. These constraints are analogous to the familiar torsion constraint $T_{ab}{}^c = 0$ commonly used in superspace, which does not delete any degrees of freedom. To be more specific, we saw in the fluctuation analysis in [9] that some components of superfield $\Delta_\alpha{}^\beta$ [9] entering in $E_\alpha \equiv E_\alpha{}^M \partial_M$ can be expressed in terms of $H_\alpha{}^b$ under (2.7):

$$E_\alpha = \Psi^{1/2}\{\exp(\tfrac{1}{2}\Delta)\}_\alpha{}^\beta (D_\beta + H_\beta{}^m \partial_m) \ , \tag{2.8a}$$

$$\Delta_\alpha{}^\beta \equiv \left[ i\Psi^a \gamma_a + \tfrac{1}{2}\Psi^{ab}\gamma_{ab} + \tfrac{i}{6}\Psi^{[3]}\gamma_{[3]} + \tfrac{1}{24}\Psi^{[4]}\gamma_{[4]} + \tfrac{i}{120}\Psi^{[5]}\gamma_{[5]} \right]_\alpha{}^\beta \ , \tag{2.8b}$$

$$X_{ab}{}^c = \tfrac{1}{16}\left[ (\gamma_{ab})^{\gamma\delta} D_\gamma H_\delta{}^c + 16\delta_{[a}{}^c \Psi_{b]} - 16\Psi_{ab}{}^c \right] \ , \tag{2.8c}$$

$$X_{a_1\cdots a_5}{}^c = \tfrac{1}{16}\left[ i(\gamma_{a_1\cdots a_5})^{\gamma\delta} D_\gamma H_\delta{}^c + \tfrac{2}{3}\delta_{[a_1}{}^c \Psi_{a_2\cdots a_5]} + \tfrac{2}{15}\epsilon^c{}_{a_1\cdots a_5}{}^{[5]}\Psi_{[5]} \right] \ . \tag{2.8d}$$

As seen from (2.8c) and (2.8d), the constraints (2.7) are just equivalent to determining $\Psi^{[1]}, \Psi^{[3]}, \Psi^{[4]}$ and $\Psi^{[5]}$ in terms only of $H_\alpha{}^b$ [9]. This situation is similar to the constraint $T_{ab}{}^c = 0$ in order to express the Lorentz connection $\phi_{ma}{}^b$ in terms of the vielbein $e_a{}^m$. Therefore we stress that there is no loss of degrees of freedom under the covariant constraints (2.7).

---

[4]In this paper, the symbol $_{[n]}$ in general denotes the total antisymmetrization of $n$ indices, *e.g.*, $(\gamma^{[5]})_{\alpha\beta}(\gamma_{[5]})_{\gamma\delta} \equiv (\gamma^{a_1\cdots a_5})_{\alpha\beta}(\gamma_{a_1\cdots a_5})_{\gamma\delta}$.

The original tensor superfields $X$'s or $U$'s are reduced into more fundamental irreducible components. For example, the original $U_{abc}$ has 605 components considering their symmetries, which can be decomposed into $U_{abc} = \mathcal{A}^{\{429\}}_{abc} + \mathcal{A}^{\{165\}}_{abc} + \eta_{a[b}\mathcal{A}^{\{11\}}_{c]}$, where $\mathcal{A}^{\{165\}}_{abc}$ is totally antisymmetric, while the **429**-part is traceless: $\mathcal{A}^{\{429\}}{}_a{}^a{}_c = 0$, as the remainder degrees of freedom are out of the original $11 \times 55 = 605$ components. In a similar fashion, we can decompose the rest of $U$'s as

$$\begin{aligned}
U_{abc} &= +\mathcal{A}^{\{429\}}_{abc} + \mathcal{A}^{\{165\}}_{abc} + \eta_{a[b}\mathcal{A}^{\{11\}}_{c]} \;, \\
U_{ab}{}^{cd} &= +\mathcal{A}^{\{2,574\}}_{ab}{}^{cd} + \mathcal{A}^{\{330\}}_{ab}{}^{cd} + \delta_{[a|}{}^{[c|}\mathcal{A}^{\{65\}|d]}_{|b]} + \delta_{[a|}{}^{[c|}\mathcal{A}^{\{55\}|d]}_{|b]} + \delta_{[a|}{}^{[c|}\delta_{|b]}{}^{|d]}\mathcal{A}^{\{1\}} \;, \\
U_{a_1\cdots a_5 bc} &= +\mathcal{A}^{\{17,160\}}_{a_1\cdots a_5 bc} + \tfrac{1}{120}\epsilon_{a_1\cdots a_5[b|}{}^{d_1\cdots d_5}\mathcal{A}^{\{4,290\}}_{d_1\cdots d_5|c]} + \eta_{[b|[a_1|}\mathcal{A}^{\{3,003\}}_{|a_2\cdots a_5]|c]} \\
&\quad + \eta_{[b|[a_1|}\mathcal{A}^{\{462\}}_{|a_2\cdots a_5]|c]} + \tfrac{1}{24}\epsilon_{a_1\cdots a_5 bc}{}^{[4]}\mathcal{B}^{\{330\}}_{[4]} + \eta_{b[a_1|}\eta_{|a_2|c}\mathcal{B}^{\{165\}}_{|a_3 a_4 a_5]} \;.
\end{aligned} \qquad (2.9)$$

Substituting each irreducible component back into (2.3), we get the set of algebraic conditions to be satisfied for each irreducible components:

$$i(\gamma^d)_{(\alpha\beta|}(\gamma_{de})_{|\gamma\delta)}\mathcal{A}^{\{1\}} = 0 \;, \qquad (2.10\text{a})$$

$$(\gamma^d)_{(\alpha\beta|}(\gamma_{[d|})_{|\gamma\delta)}\mathcal{A}^{\{11\}}_{|e]} = 0 \;, \qquad (2.10\text{b})$$

$$i(\gamma^a)_{(\alpha\beta|}(\gamma^{bc})_{|\gamma\delta)}\left[\eta_{a[b|}\mathcal{A}^{\{55\}}_{|c]e} - \eta_{e[b|}\mathcal{A}^{\{55\}}_{|c]a}\right] = 0 \;, \qquad (2.10\text{c})$$

$$i(\gamma^a)_{(\alpha\beta|}(\gamma^{bc})_{|\gamma\delta)}\left[\eta_{a[b|}\mathcal{A}^{\{65\}}_{|c]e} - \eta_{e[b|}\mathcal{A}^{\{65\}}_{|c]a}\right] = 0 \;, \qquad (2.10\text{d})$$

$$(\gamma^d)_{(\alpha\beta|}(\gamma_{de}{}^{[3]})_{|\gamma\delta)}\mathcal{B}^{\{165\}}_{[3]} = 0 \;, \qquad (2.10\text{e})$$

$$\tfrac{i}{2}(\gamma^d)_{(\alpha\beta|}(\gamma^{ab})_{|\gamma\delta)}\mathcal{A}^{\{330\}}_{abde} - \tfrac{1}{24}\tfrac{1}{120}(\gamma^d)_{(\alpha\beta|}(\gamma^{[5]})_{|\gamma\delta)}\,\epsilon_{[5]de}{}^{[4]}\mathcal{B}^{\{330\}}_{[4]} = 0 \;, \qquad (2.10\text{f})$$

$$-(\gamma^a)_{(\alpha\beta|}(\gamma^b)_{|\gamma\delta)}\mathcal{A}^{\{429\}}_{abe} + \tfrac{1}{4}(\gamma^{ab})_{(\alpha\beta|}(\gamma_{de})_{|\gamma\delta)}X_{ab}{}^d = 0 \;, \qquad (2.10\text{g})$$

$$(\gamma^d)_{(\alpha\beta|}(\gamma^{b[4]})_{|\gamma\delta)}\eta_{[d|b}\mathcal{A}^{\{462\}}_{[4]|e]} = 0 \;, \qquad (2.10\text{h})$$

$$i(\gamma^d)_{(\alpha\beta|}(\gamma^{ab})_{|\gamma\delta)}\mathcal{A}^{\{2,574\}}_{abde} = 0 \;, \qquad (2.10\text{i})$$

$$(\gamma^d)_{(\alpha\beta|}(\gamma^{b[4]})_{|\gamma\delta)}\eta_{[d|b}\mathcal{A}^{\{3,003\}}_{[4]e} = 0 \;, \qquad (2.10\text{j})$$

$$\tfrac{1}{120}(\gamma^d)_{(\alpha\beta|}(\gamma^{[5]})_{|\gamma\delta)}\epsilon_{[5][d|}{}^{[5]'}\mathcal{A}^{\{4,290\}}_{[5]'|e]} + \tfrac{i}{2}(\gamma^{[5]})_{(\alpha\beta|}(\gamma_{de})_{|\gamma\delta)}X_{[5]}{}^d = 0 \;, \qquad (2.10\text{k})$$

$$(\gamma^d)_{(\alpha\beta|}(\gamma^{[5]})_{|\gamma\delta)}\mathcal{A}^{\{17,160\}}_{[5]de} = 0 \;. \qquad (2.10\ell)$$

The conceptually important ingredient here is that different irreducible components in (2.10), such as **165** vs. **2,574** will not interfere with each other. Moreover, $\mathcal{A}^{\{330\}}_{abcd}$ and $\mathcal{B}^{\{330\}}_{abcd}$ in (2.10f), $\mathcal{A}^{\{429\}}_{abc}$ and $X_{ab}{}^c$ in (2.10g), or $\mathcal{A}^{\{4,290\}}_{[5]a}$ and $X_{[5]}{}^c$ in (2.10k) are proportional to each other.

From now on, we use heavily the Fierz-type identities (5.1) - (5.4) and Lemmas (5.5) - (5.8) that will be given separately in section 6. We start with eq. (2.10a). This condition is identically satisfied for arbitrary $\mathcal{A}^{\{1\}}$ due to the well-known Fierz identity in 11D

$$(\gamma^{ab})_{(\alpha\beta|}(\gamma_b)_{|\gamma\delta)} \equiv 0 \ . \tag{2.11}$$

Next, due to Lemma 2 in (5.5), eq. (2.10b) implies that

$$\mathcal{A}_a^{\{11\}} = 0 \ . \tag{2.12}$$

Eqs. (2.10c) and (2.10d) satisfy the assumption of Lemma 2, when $B_{a,bc}$ is identified with

$$B_{a,bc} \ \rightarrow \ \eta_{ac}\mathcal{A}_{be}^{\{n\}} - \eta_{ab}\mathcal{A}_{ce}^{\{n\}} - \eta_{ec}\mathcal{A}_{ba}^{\{n\}} + \eta_{eb}\mathcal{A}_{ca}^{\{n\}} \tag{2.13}$$

both for $n = 55$ and $n = 66$. Here we use the 'arrow' symbol instead of 'equality', due to the free index $e$, while $B_{a,bc}$ on the l.h.s. can be arbitrary including any such 'free' index. Therefore corresponding to (5.6a), we get the condition

$$\eta_{a[b}\mathcal{A}_{c]e}^{\{n\}} - 8\eta_{e[b}\mathcal{A}_{c]a}^{\{n\}} + \eta_{ea}\mathcal{A}_{[bc]}^{\{n\}} - \eta_{e[b|}(\mathcal{A}_{a|c]}^{\{n\}} + \mathcal{A}_{|c]a}^{\{n\}}) = 0 \ , \tag{2.14}$$

both for $n = 55$ and $n = 66$. Now the $ae$- and $ce$-contractions respectively yield

$$\mathcal{A}_{bc}^{\{55\}} = 0 \ , \quad \mathcal{A}_{bc}^{\{66\}} = 0 \ . \tag{2.15}$$

Eqs. (2.10e), (2.10h), (2.10j) and (2.10$\ell$) are solved based on Lemma 3 in (5.8), namely they imply nothing other than the vanishing of

$$\mathcal{B}_{[3]}^{\{165\}} = 0 \ , \quad \mathcal{A}_{[4]a}^{\{462\}} = 0 \ , \quad \mathcal{A}_{[4]a}^{\{3,003\}} = 0 \ , \quad \mathcal{A}_{[5]ab}^{\{17,160\}} = 0 \ . \tag{2.16}$$

Eq. (2.10f) is understood as a sum of (5.2) + (5.4), when $B_{a,bc}$ and $B_{a,[5]}$ are identified with

$$B_{a,bc} \ \rightarrow \ 24 A_{bcae}^{\{330\}} \ , \quad B_{a,[5]} \ \rightarrow \ -\epsilon_{[5]ae}{}^{[4]} \mathcal{B}_{[4]}^{\{330\}} \ . \tag{2.17}$$

The like terms of the types $(\gamma^a)_{\alpha\beta}(\gamma^{[5]})_{\gamma\delta}$ and $(\gamma^{[2]})_{\alpha\beta}(\gamma^{[2]'})_{\gamma\delta}$ in the sum (5.2) + (5.4) yield respectively the conditions

$$9B_{a,bc_1\cdots c_4} - \tfrac{1}{24}\eta_{a[b|}B_{d,}{}^d{}_{|c_1\cdots c_4]} + \tfrac{1}{24}B_{[b|,a|c_1\cdots c_4]} = 0 \ , \tag{2.18a}$$

$$B_{a,}{}^a{}_{[4]} = 0 \ , \tag{2.18b}$$

both of which have contributions only from $B_{a,[5]}$. Eq. (2.18b) deletes the middle term in (2.18a). Other non-trivial like terms are of the type $(\gamma^{[2]})_{\alpha\beta}(\gamma^{[6]})_{\gamma\delta}$, which have contributions both from the $B_{a,bc}$ and $B_{a,[4]}$ terms, yielding the condition

$$\tfrac{5}{2}(\delta_a{}^c\delta_b{}^d - \delta_b{}^c\delta_a{}^d)B_{[f,gh]} - \tfrac{1}{30}\epsilon_b{}^{cde[4]}{}_{fgh}B_{a,e[4]} = 0 \ . \tag{2.19}$$

Now if we look into only the $[abcd]$-component of this equation, and multiply it by $\epsilon_{cd}{}^{abghk_1\cdots k_5}$, we get the condition

$$B_{f,k_1\cdots k_5} = 0 \ , \tag{2.20}$$

up to terms that vanish upon using eq. (2.18b). This implies consistently with (2.18) that

$$B_{a,b_1\cdots b_5} = 0 \implies \mathcal{B}^{\{330\}}_{[4]} = 0 \ . \tag{2.21}$$

Now once $\mathcal{B}^{\{330\}}$'s does not contribute, then only the first term in (2.10f) remains, which in turn implies *via* (5.7) in Lemma 2 that

$$\mathcal{A}^{\{330\}}_{abcd} = 0 \ , \tag{2.22}$$

because $\mathcal{A}^{\{330\}}{}_{ac}{}^a{}_d = 0$ manifestly, upon the identification $B_{a,bc} \to \mathcal{A}^{\{330\}}_{bcad}$.

Eq. (2.10i) has the $\gamma$-matrix structure of (5.2) with $B_{a,bc}$ identified with

$$B_{a,bc} \to \mathcal{A}^{\{2,574\}}_{bcea} \ , \quad B_{a,}{}^a{}_c \to \mathcal{A}^{\{2,574\}}{}^a{}_{ace} = 0 \ , \tag{2.23}$$

so that the assumption of (5.7) in Lemma 2 is satisfied, and therefore $B_{a,bc} = 0$, *i.e.*,

$$\mathcal{A}^{\{2,574\}}_{bcea} = 0 \ . \tag{2.24}$$

At this stage, eqs. (2.10g) and (2.10k) are the only remaining conditions to be solved. Eq. (2.10g) is regarded as the sum $(5.1) + (5.3)$, when $A_{a,b}$ for the former, and $A_{ab,cd}$ for the latter are respectively identified with

$$A_{a,b} \to -24\mathcal{A}^{\{429\}}_{abe} \ , \quad A_{ab,}{}^{cd} \to -6\eta_{e[a|}X^{cd}{}_{|b]} - 6\delta_e{}^{[c}X_{ab}{}^{d]} \ . \tag{2.25}$$

Accordingly, it is convenient to rewrite this $(5.1) + (5.3)$ in terms of $A_{a,b}$ and $A_{ab,cd}$ instead of $\mathcal{A}$'s:

$$\frac{1}{24}(\gamma^a)_{(\alpha\beta|}(\gamma^b)_{|\gamma\delta)}A_{a,b} + \frac{1}{96}(\gamma^{ab})_{(\alpha\beta|}(\gamma^{cd})_{|\gamma\delta)}A_{ab,cd}$$
$$= +\frac{1}{48}(\gamma^a)_{\alpha\beta}(\gamma^b)_{\gamma\delta}(18A_{a,b} - \eta_{ab}A_{c,}{}^c + 2A_{ac,b}{}^c)$$
$$+ \frac{1}{96}(\gamma^{ab})_{\alpha\beta}(\gamma_{cd})_{\gamma\delta}\Big[ +\delta_{[a|}{}^{[c|}A_{|b]}{}^{|d]} - \frac{1}{2}\delta_{[a}{}^c\delta_{b]}{}^d A_{f,}{}^f$$
$$+ 9A_{ab,}{}^{cd} + \frac{1}{2}A_{[a|}{}^{[c|}{}_{,|b]}{}^{|d]} - \delta_{[a|}{}^{[c|}A_{|b]f,}{}^{|d]f} \Big]$$
$$+ \frac{1}{5,760}(\gamma^{[3]ab})_{\alpha\beta}(\gamma_{[3]}{}^{cd})_{\gamma\delta}\Big[ +\frac{5}{2}\delta_{[a}{}^{[c}A_{b],}{}^{d]} - \frac{1}{2}\delta_{[a}{}^c\delta_{b]}{}^d A_{f,}{}^f$$
$$- 10A_{ab,}{}^{cd} - 5A_{[a|}{}^{[c|}{}_{,|b]}{}^{|d]} + \frac{5}{2}\delta_{[a|}{}^{[c|}A_{|b]f,}{}^{|d]f} \Big] = 0 \ . \tag{2.26}$$

Since each of the different $\gamma$-matrix structure is independent, we have the following three conditions

$$18A_{a,b} - \eta_{ab}A_{c,}{}^c + 2A_{ac,b}{}^c = 0 \ , \tag{2.27a}$$

$$\delta_{[a|}{}^{[c|}A_{|b]}{}^{|d]} - \frac{1}{2}\delta_{[a}{}^c\delta_{b]}{}^d A_{f,}{}^f + 9A_{ab,cd} + \frac{1}{2}A_{[a|}{}^{[c|}{}_{,|b]}{}^{|d]} - \delta_{[a|}{}^{[c|}A_{|b]f,}{}^{|d]f} = 0 \ , \tag{2.27b}$$

$$\frac{5}{2}\delta_{[a}{}^{[c}A_{b],}{}^{d]} - \frac{1}{2}\delta_{[a}{}^c\delta_{b]}{}^d A_{f,}{}^f - 10A_{ab,cd} - 5A_{[a|}{}^{[c|}{}_{,|b]}{}^{|d]} + \frac{5}{2}\delta_{[a|}{}^{[c|}A_{|b]f,}{}^{|d]f} = 0 \ . \tag{2.27c}$$

Obviously, (2.27a) implies that $A_{a,}{}^a = 0$, already satisfied by the tracelessness of $\mathcal{A}^{\{429\}}_{abc}$. Using this back in (2.27a) implies that

$$9A_{a,b} + A_{ac,b}{}^c = 0 \ . \tag{2.28}$$

On the other hand, by contracting the $bd$-indices in (2.27b), we get

$$9A_{a,}{}^c - 7A_{ab,}{}^b = 0 \ . \tag{2.29}$$

Obviously, (2.29) and (2.28) lead to $A_{ab} = 0$, $A_{ab}{}^{cb} = 0$, which *via* (2.27b) and (2.27c) implies also that $A_{abcd} = 0$. Therefore we get

$$A_{a,b} = 0 \implies \mathcal{A}_{abc}^{\{429\}} = 0 \ , \tag{2.30a}$$

$$A_{ab,cd} = 0 \implies X_{ab}{}^c = 0 \ . \tag{2.30b}$$

We are now left with the condition (2.10k). As mentioned before, we can regard $\mathcal{A}_{[5]a}^{\{4,290\}}$ and $X_{[5]}{}^c$ as proportional to each other: $\mathcal{A}_{[5]a}^{\{4,290\}} = \text{const.}\ X_{[5]a}$. Then finding a non-trivial solution for (2.10k) is equivalent to deducing a non-trivial solution for the unknown parameters $a$ and $b$ other than $a = b = 0$ in the equality

$$ia(\gamma^d)_{(\alpha\beta|}(\gamma_{[d|}{}^{[5]})_{|\gamma\delta)}X_{[5]|e]} + ib(\gamma^{[5]})_{(\alpha\beta|}(\gamma_{de})_{|\gamma\delta)}X_{[5]}{}^d = 0 \ , \tag{2.31}$$

for an arbitrary $X_{[5]}{}^c \neq 0$. The simplest way to get more explicit conditions from (2.31) is to multiply it by $(\gamma^{ab})^{\alpha\beta}$, and contract the indices $\alpha\beta$:

$$(\gamma^{ab})^{\alpha\beta} [\text{ LHS of } (2.31)_{\alpha\beta\gamma\delta}]$$
$$= +20i(3a + b)(\gamma^{[a|[4]})_{\gamma\delta}X_{[4]}{}^{|b]}{}_e - 8i(a + 5b)\delta_e{}^{[a|}(\gamma^{[5]})_{\gamma\delta}X_{[5]}{}^{|b]}$$
$$+ 40i(a - b)(\gamma_e{}^{[4]})_{\gamma\delta}X_{[4]}{}^{[ab]} + 40ib(\gamma^{[4][a|})_{\gamma\delta}X_{[4]e}{}^{|b]} = 0 \ . \tag{2.32}$$

We next multiply (2.32) by $(\gamma^b)_\epsilon{}^\gamma$, to get the only solutions $a = b = 0$ as

$$(60a + 28b)(\gamma^{a[5]})_{\epsilon\delta}X_{[5]e} + (40a + 8b)(\gamma^{e[5]})_{\epsilon\delta}X_{[5]a} + (320a + 160b)(\gamma^{[4]})_{\epsilon\delta}X_{[4]}{}^{ae} = 0$$

$$\implies a = b = 0 \ . \tag{2.33}$$

This is because the $\gamma^{[4]}$-term yields $2a + b = 0$, while the multiplication of the $\gamma^{[6]}$-terms by $i\gamma^a$ yields $25a + 11b = 0$. Therefore the only solutions to the condition (2.10k) are

$$\mathcal{A}_{[5]a}^{\{4,290\}} = 0 \ , \quad X_{[5]}{}^c = 0 \ . \tag{2.34}$$

Collecting all the results above, *i.e.*, (2.12), (2.15), (2.16), (2.21), (2.22), (2.24), (2.30), and (2.34), we reach the conclusion that among all the components of $U$'s entering (2.5), *except for* the singlet component $\mathcal{A}^{\{1\}}$ in (2.9), as well as all the $X$'s in (2.4), should be zero, in order to satisfy the F-BI (2.1) at $d = 0$ under the conditions (2.2). Therefore, the only possible form for $F_{\alpha\beta cd}$ is

$$F_{\alpha\beta cd} = +2(\gamma_{cd})_{\alpha\beta}\mathcal{A}^{\{1\}} + i(\gamma^b)_{\alpha\beta}\mathcal{A}_{bcd}^{\{165\}} \ . \tag{2.35}$$

The remaining component $\mathcal{A}_{abc}^{\{165\}}$ here is due to the fact that this component does *not* enter any of the conditions in (2.10). In other words, we can still have a term proportional to $\mathcal{A}_{abc}^{\{165\}}$ in (2.35).

However, we point out some degrees of freedom of superfield redefinition of the potential superfield $A_{ABC}$. This is associated with the definition of the superfield strength $F_{ABCD}$. In fact, consider the shift[5]

$$A_{abc} \rightarrow A_{abc} + \mathcal{A}_{abc}^{\{165\}} \quad , \tag{2.36}$$

keeping other components among $A_{ABC}$ intact. This can absorb the $\mathcal{A}^{\{165\}}$-term in (2.35), while any of the constraints at $d < 0$ in (2.2) are maintained. For example for $F_{\alpha\beta\gamma\delta}$, we have

$$F_{\alpha\beta\gamma\delta} \equiv \tfrac{1}{6}\nabla_{(\alpha}A_{\beta\gamma\delta)} - \tfrac{1}{4}T_{(\alpha\beta|}{}^\epsilon A_{\epsilon|\gamma\delta)} - \tfrac{1}{4}T_{(\alpha\beta|}{}^e A_{e|\gamma\delta)} \quad , \tag{2.37}$$

which is intact under the shift (2.36). The same is also true for $F_{\alpha\beta\gamma d}$, which we skip here.

Based on these results and considerations, we conclude that the only degree of freedom possibly embedding M-theory corrections for the superspace constraints at $d = 0$ is the singlet component $\mathcal{A}^{\{1\}}$ in (2.35), which implies that $F_{\alpha\beta cd}$ can be only of the form

$$F_{\alpha\beta cd} = \tfrac{1}{2}(\gamma_{cd})_{\alpha\beta}\, e^\Phi \tag{2.38}$$

proportional to the on-shell physical superfield constraint [11], scaled by some real scalar superfield $\Phi$.

We mention here the importance of the conventional constraints (2.7). If these constraints *were not* imposed, we would have such corrections as

$$T_{\alpha\beta}{}^c = i(\gamma^c)_{\alpha\beta} + i(M\gamma^c)_{(\alpha\beta)} \quad , \quad F_{\alpha\beta cd} = \tfrac{1}{2}(\gamma_{cd})_{\alpha\beta} + \tfrac{1}{2}(M\gamma_{cd})_{(\alpha\beta)} \quad , \tag{2.39}$$

for an *arbitrary* $32 \times 32$ matrix $M_\alpha{}^\beta$ satisfying (2.3) at the linear order in $M$. However, these corrections can not embed M-theory corrections, because they contribute as redundant degrees of freedom. We can also show that the gravitational superfield equation will not be modified at the linear order in $M$, because the $(\alpha\beta c, \delta)$-type BI is not modified by $M$. Therefore, such a matrix $M$ is not enough for embedding M-theory corrections. We skip the details here, leaving them for a future publication.

Our analysis so far is concerned only with the linear order terms in (2.3) in the fluctuations in $F_{\alpha\beta cd}$ and $T_{\alpha\beta}{}^c$ for M-theory corrections. However, even if we include the bilinear-order terms in (2.3), our conclusion above remains intact. In other words, the linear-order satisfaction of the $F$-BI at $d = 0$ only by the limited form (2.38) will not be affected by the inclusion of next bilinear order terms. To put it differently, once the most general correction of the $F_{\alpha\beta cd}$ has determined as in (2.3) at the linear order, in particular with no corrections for $T_{\alpha\beta}{}^c$, then it is straightforward to confirm that (2.38) satisfies also the $F$-BI at $d = 0$ to 'all orders' in the expansion in terms of $\Phi$, because the only correction is just a scaling of $F_{\alpha\beta cd}$, while there is no derivative involved in the $d = 0$ BI (2.3).

This 'no-go theorem' established here is not so surprising from the viewpoint that (2.11) is the only available Fierz identity of the type $(\gamma^{[m]})_{(\alpha\beta}(\gamma^{[n]})_{|\gamma\delta)} = 0$ allowing arbitrary integers $m$ and $n$. In other words, such conditions as (2.3) has too many free indices to allow more degrees of freedom other than the singlet component $\mathcal{A}^{\{1\}}$ as in (2.10a).

---
[5]Notice the crucial difference of the symbol $A$'s from $\mathcal{A}$'s which should not be confused with the former.

## 3. Effect of Scalar Superfield $\Phi$

We have seen that the only possible correction of superspace constraints at $d=0$ is the scaling of $F_{\alpha\beta cd}$ in (2.38) by a real scalar superfield $\Phi$. The next question is what sort of M-theory corrections can be embedded into this scalar superfield $\Phi$.

The answer can be easily deduced from dimensional considerations. First, since the scalar superfield $\Phi$ is at $d=0$, its spinorial derivative enters into the $d=1/2$ constraints $T_{\alpha\beta}{}^\gamma$, $T_{\alpha b}{}^c$ and $F_{\alpha bcd}$. Therefore we have again the spinorial superfield $J_\alpha$ [6] related to $\Phi$ by[6]

$$\nabla_\alpha \Phi \equiv \xi J_\alpha ~, \tag{3.1}$$

with an appropriate constant $\xi$ like in refs. [9][6]. The most general forms for these constraints are now

$$T_{\alpha b}{}^c = \alpha_1 \delta_b{}^c J_\alpha + \alpha_2 (\gamma_b{}^c)_\alpha{}^\beta J_\beta ~, \tag{3.2a}$$

$$F_{\alpha bcd} = i\eta e^\Phi (\gamma_{bcd})_\alpha{}^\beta J_\beta ~, \tag{3.2b}$$

$$T_{\alpha\beta}{}^\gamma = \beta_1 \delta_{(\alpha}{}^\gamma J_{\beta)} + \beta_2 (\gamma^a)_{\alpha\beta}(\gamma_a)^{\gamma\delta} J_\delta + \beta_3 (\gamma^{[2]})_{\alpha\beta}(\gamma_{[2]})^{\gamma\delta} J_\delta ~, \tag{3.2c}$$

with the unknown coefficients $\alpha_1$, $\alpha_2$, $\eta$, $\beta_1$, $\beta_2$, $\beta_3$. The satisfaction of $d=1/2$ BIs yields the following relationships among the coefficients

$$\beta_1 = -\tfrac{5}{2}(\alpha_1+\alpha_2) + \xi ~, \quad \beta_2 = -\tfrac{1}{4}(9\alpha_1+5\alpha_2) + \tfrac{3}{4}\xi ~,$$
$$\beta_3 = +\tfrac{3}{8}(\alpha_1+\alpha_2) - \tfrac{1}{8}\xi ~, \quad \eta = +\tfrac{3}{2}\alpha_1 - \tfrac{1}{2}\xi ~. \tag{3.3}$$

Note the important fact that the exponential function $e^\Phi$ is needed in $F_{\alpha\beta\gamma d}$, while such a factor is absent in $T_{\alpha b}{}^c$ and $T_{\alpha\beta}{}^\gamma$ in order to satisfy the $F$-BI at $d=1/2$. For example, the $(\alpha\beta cde)$-type $F$-BI tells us that all the $F_{abcd}$-linear terms in $T_{\alpha b}{}^\gamma$ stay the same with *no* exponential function $e^\Phi$:

$$T_{\alpha b}{}^\gamma|_F = \tfrac{i}{144}(\gamma_b{}^{[4]} F_{[4]} + 8\gamma^{[3]} F_{b[3]})_\alpha{}^\gamma ~, \tag{3.4}$$

where the symbol $|_F$ denotes the $F_{abcd}$-linear part of $T_{\alpha b}{}^\gamma$. As is well-known, the $(abcde)$-type $F$-BI at $d=3/2$ gives the expression for $\nabla_\alpha F_{bcde}$ in terms of $T_{ab}{}^\gamma$ and $F_{\alpha bcd}$, where the latter contains the linear $J$'s as

$$\nabla_\alpha F_{bcde}|_{\nabla J} = \tfrac{i}{6}\eta e^\Phi (\gamma_{[bcd|})_\alpha{}^\beta \nabla_{|e]} J_\beta ~, \tag{3.5}$$

Here the factor $e^\Phi$ is involved *via* (3.2b). On the other hand, the $(\alpha bcd)$-type $T$-BI gives the relationship

$$R_{\alpha bcd}|_{\nabla J} = \alpha_1 \eta_{b[c} \nabla_{d]} J_\alpha + \alpha_2 (\gamma_{cd})_\alpha{}^\beta \nabla_b J_\beta ~, \tag{3.6}$$

with *no* exponential factor $e^\Phi$ with the $J$'s. When (3.5) and (3.6) are used in the $(a\beta\gamma,\delta)$-type BI, the former produces an exponential factor $e^\Phi$, while the latter does *not*.

---

[6]We do not take the standpoint in [7] that there is no auxiliary spinorial superfield in $d=1/2$ constraints at least temporarily, for the sake of argument here.

In order to avoid this mismatch, we are forced to put $\eta = 0$, which in turn *via* (3.3) implies that

$$\eta = 0 \ , \quad \alpha_1 = \tfrac{1}{3}\xi \ , \quad \beta_1 = -\tfrac{5}{2}\alpha_2 + \tfrac{1}{6}\xi \ , \quad \beta_2 = -\tfrac{5}{4}\alpha_2 \ , \quad \beta_3 = +\tfrac{3}{8}\alpha_2 \ . \quad (3.7)$$

The gravitino superfield equation can be obtained from the $(\alpha\beta\gamma,\delta)$-type BI, by contracting spinorial indices in several different ways, which should be consistent with each other. One way is to multiply this BI by $i(\gamma^a)_\alpha{}^\beta \delta_\delta{}^\gamma$ to get the trace part of the gravitino superfield equation

$$- \tfrac{185}{8}(\gamma^{ab})_{\alpha\beta} T_{ab}{}^\beta + \tfrac{23i}{6}\xi(\gamma^a)_{\alpha\beta} \nabla_a J^\beta = 0 \ , \quad (3.8)$$

while another way is to multiply the same BI by $i(\gamma^a)_\delta{}^\gamma$ to get

$$\tfrac{15}{8}(\gamma^{ab})_{\alpha\beta} T_{ab}{}^\beta + \tfrac{11i}{6}\xi(\gamma^a)_{\alpha\beta} \nabla_a J^\beta = 0 \ , \quad (3.9)$$

up to terms, such as $J^2$ or the $J$'s with fundamental physical superfields ignored as higher orders. Note that all the $\alpha_2$-dependent terms cancel each other in these equations. obviously, (3.8) and (3.9) lead to the conclusion that $\xi = 0$ as the only possible solution. Unfortunately, this is a trivial solution, because this implies that $J_\alpha = 0$ in (3.1), so that $\Phi =$ const. Even though we did not mention, there are also other additional conditions on the independent parameters in (3.7), that will not change the conclusion here. This is because they provide *more* stronger conditions on the unknown parameters, but they never avoid the conclusion $\xi = 0$ above. Note also that our result here is in agreement with the argument about the absence of the off-shell $J$-superfield in [7].

From these considerations, we conclude that the scalar superfield $\Phi$ embedded into $F_{\alpha\beta cd}$ as the exponent above is *not* enough to embed M-theory corrections, as long as the $F$-BIs are not modified by Chern-Simons terms.

## 4. Fermionic $\kappa$-Symmetry and Chern-Simons Modification

We next consider the fermionic $\kappa$-symmetry of supermembrane action [12], which justifies our assumption (i) about the vanishing $F$-constraints at $d < 0$. The standard supermembrane action is [12]

$$I \equiv \int d^3\sigma \Big[ + \tfrac{1}{2}\sqrt{-g} g^{ij} \eta_{ab} \Pi_i{}^a \Pi_j{}^b - \tfrac{1}{2}\sqrt{-g} - \tfrac{1}{3}\epsilon^{ijk}\Pi_i{}^A \Pi_j{}^B \Pi_k{}^C A_{CBA} \Big] \ . \quad (4.1)$$

with the pull-backs $\Pi_i{}^A \equiv (\partial_i Z^M) E_M{}^A$, for the superspace coordinates $Z^M$ and the inverse vielbein $E_M{}^A$ in the 11D superspace we are dealing with. The fermionic $\kappa$-symmetry is dictated by [12]

$$\delta_\kappa E^\alpha = (I+\Gamma)^\alpha{}_\beta \kappa^\beta \ , \quad \delta_\kappa E^a = 0 \ , \quad \Gamma^{\alpha\beta} \equiv \tfrac{i}{6}\epsilon^{ijk}\Pi_i{}^a \Pi_j{}^b \Pi_k{}^c (\gamma_{abc})^{\alpha\beta} \ , \quad (4.2)$$

where $\delta_\kappa E^A \equiv (\delta_\kappa Z^M) E_M{}^A$. The general variation formula under $\delta_\kappa E^a = 0$ is

$$\delta_\kappa I = \sqrt{-g} g^{ij} (\delta_\kappa E^\alpha) \Pi_i{}^B T_{B\alpha}{}^d \Pi_{jd} + \tfrac{1}{3}\epsilon^{ijk}(\delta_\kappa E^\alpha) \Pi_i{}^B \Pi_j{}^C \Pi_k{}^D F_{DCB\alpha} \ . \quad (4.3)$$

The first term is from the variation of the kinetic term, while the second one is from the Wess-Zumino-Novikov-Witten (WZNW) term in (4.1).

We now study whether the non-zero $F$-constraints $F_{\alpha\beta\gamma\delta} \neq 0$, $F_{\alpha\beta\gamma d} \neq 0$ at $d < 0$ are compatible with this fermionic symmetry (4.2). Using (4.3), we easily see that if there are such non-trivial $F$-constraints at $d < 0$, they will contribute only to the variation of the WZNW-term in (4.3), that are *not* simply cancelled by the variation of the kinetic term:

$$\delta_\kappa I = +\tfrac{1}{3}\epsilon^{ijk}(\delta_\kappa E^\alpha)\Pi_i{}^\beta \Pi_j{}^\gamma \Pi_k{}^\delta F_{\delta\gamma\beta\alpha} + \epsilon^{ijk}(\delta_\kappa E^\alpha)\Pi_i{}^\beta \Pi_j{}^\gamma \Pi_k{}^d F_{d\gamma\beta\alpha} \ . \tag{4.4}$$

It is unlikely that new corrections due to other radical and non-conventional corrections such as $\delta_\kappa E^a = 0$ itself, or due to the addition of some other terms to the action $I$ itself can lead to the fermionic $\kappa$-invariance of the conventional supermembrane action (4.1), because such corrections occur only at $d < 0$, which do not seem to communicate with other $d \geq 0$ constraints.[7]

## 5. Useful Algebraic Lemmas

In what follows, we list up some useful algebraic lemmas and relationships that play decisive roles for our conclusions in this paper. We start with the Fierz identity

$$\begin{aligned}
\tfrac{1}{24}(\gamma^a)_{(\alpha\beta|}(\gamma^b)_{|\gamma\delta)} A_{a,b} =& +(\gamma^a)_{\alpha\beta}(\gamma^b)_{\gamma\delta}\Big(\tfrac{3}{8}A_{a,b} - \tfrac{1}{48}\eta_{ab}A_c{}^c\Big) \\
& +(\gamma_{ab})_{\alpha\beta}(\gamma^{cd})_{\gamma\delta}\Big[\tfrac{1}{96}\delta_{[c|}{}^{[a|}A_{|d],}{}^{|b]} - \tfrac{1}{192}\delta_{[c}{}^a\delta_{d]}{}^b A_c{}^c\Big] \\
& +(\gamma^{[4]a})_{\alpha\beta}(\gamma_{[4]}{}^b)_{\gamma\delta}\Big(\tfrac{1}{576}A_{a,b} - \tfrac{1}{5{,}760}\eta_{ab}A_c{}^c\Big) \ ,
\end{aligned} \tag{5.1}$$

where $A_{a,b}$ are any arbitrary symmetric tensor superfield. In a similar fashion for an arbitrary tensor superfield $B_{a,bc}$ with the property $B_{a,bc} = -B_{a,cb}$, we have

$$\begin{aligned}
\tfrac{i}{48}(\gamma^a)_{(\alpha\beta|}(\gamma^{bc})_{|\gamma\delta)} B_{a,bc} & \\
=& +\tfrac{i}{96}\Big[(\gamma^a)_{\alpha\beta}(\gamma^{bc})_{\gamma\delta} + (\gamma^{bc})_{\alpha\beta}(\gamma^a)_{\gamma\delta}\Big]\Big(9B_{a,bc} - B_{b,ca} - B_{c,ab} - \eta_{a[b|}B_{d,}{}^d{}_{|c]}\Big) \\
& +\tfrac{i}{2{,}304}(\gamma^{[4]a})_{\alpha\beta}(\gamma_{[4]}{}^{bc})_{\gamma\delta}\Big(B_{a,bc} - B_{b,ca} - B_{c,ab} - \tfrac{1}{5}\eta_{a[b|}B_{d,}{}^d{}_{|c]}\Big) \\
& +\tfrac{i}{1{,}152}\Big[(\gamma^{[2]})_{\alpha\beta}(\gamma_{[2]}{}^{abc})_{\gamma\delta} + (\gamma^{[2]abc})_{\alpha\beta}(\gamma_{[2]})_{\gamma\delta}\Big]B_{[a,bc]} \ .
\end{aligned} \tag{5.2}$$

Similarly, for any arbitrary tensor superfields with the properties $A_{ab,cd} = +A_{cd,ab} = (1/4)A_{[ab],[cd]}$, $A_{[ab,cd]} = 0$, $A_{ab,}{}^{ab} = 0$, we have

$$\begin{aligned}
\tfrac{1}{96}(\gamma^{ab})_{(\alpha\beta|}(\gamma^{cd})_{|\gamma\delta)} A_{ab,cd} & \\
=& +\tfrac{1}{24}(\gamma^a)_{\alpha\beta}(\gamma^b)_{\gamma\delta} A_{ac,b}{}^c \\
& +\tfrac{1}{96}(\gamma^{ab})_{\alpha\beta}(\gamma^{cd})_{\gamma\delta}\Big[9A_{ab,}{}^{cd} + A_{[a|}{}^c{}_{,|b]}{}^d - \eta_{[a|}{}^{[c|}A_{|b]e,}{}^{|d]e}\Big] \\
& +\tfrac{1}{2{,}304}(\gamma^{[3]ab})_{\alpha\beta}(\gamma_{[3]cd})_{\gamma\delta}\Big[-4A_{ab,}{}^{cd} - 4A_{[a|}{}^c{}_{,|b]}{}^d + \delta_{[a|}{}^{[c|}A_{|b]e,}{}^{|d]e}\Big] \ .
\end{aligned} \tag{5.3}$$

---

[7] One can, of course, give up such fermionic $\kappa$-symmetry of the supermembrane action (4.1) entirely, but we do not argue about the 'legitimacy' of fermionic symmetry itself in this paper.

For an arbitrary tensor superfield $B_{a,b_1\cdots b_5}$ with the totally antisymmetric indices $b_1\cdots b_5$, we have

$$\frac{1}{2,880}(\gamma^a)_{(\alpha\beta|}(\gamma^{[5]})_{|\gamma\delta)}B_{a,[5]}$$
$$= +\frac{1}{5,760}(\gamma^a)_{\alpha\beta}(\gamma^{bc_1\cdots c_4})_{\gamma\delta}\left[9B_{a,bc_1\cdots c_4} - \frac{1}{24}\eta_{a[b|}B_{d,}{}^d{}_{|c_1\cdots c_4]} + \frac{1}{24}B_{[b|,a|c_1\cdots c_4]}\right]$$
$$- \frac{1}{5,760}(\gamma^{ab})_{\alpha\beta}(\gamma_b{}^{cd_1\cdots d_4})_{\gamma\delta}\left[B_{a,cd_1\cdots d_4} - \frac{1}{300}\eta_{a[c|}B_{e,}{}^e{}_{|d_1\cdots d_4]} + \frac{1}{24}B_{[c|,a|d_1\cdots d_4]}\right]$$
$$+ \frac{1}{192}(\gamma^{bc})_{\alpha\beta}(\gamma^{[2]})_{\gamma\delta}B_{a,}{}^a{}_{bc[2]} + [(\gamma^{[5]})_{\alpha\beta}\text{-terms}] \ , \tag{5.4}$$

where we have omitted the terms with the structure of $(\gamma^{[5]})_{\alpha\beta}$, because they are independent from the terms explicitly given here, and moreover, they are too messy whose structures are not decisive for our lemma below.

Using (5.1) - (5.4), we can get the following important lemmas:

Lemma 1: *If the l.h.s. of (5.1) vanishes, then it follows that* $A_{a,b} = 0$. $\tag{5.5}$

Lemma 2: *The vanishing of the l.h.s. of (5.2) implies that the following two conditions hold:*

$$9B_{a,bc} - B_{b,ca} - B_{c,ab} - \eta_{a[b|}B_{d,}{}^d{}_{|c]} = 0 \ , \tag{5.6a}$$
$$B_{a,bc} + B_{b,ca} + B_{c,ab} = 0 \ . \tag{5.6b}$$

*In particular, when* $B_{a,}{}^a{}_b = 0$, *it follows that* $B_{a,bc} = 0$. $\tag{5.7}$

Lemma 3: *The vanishing of the l.h.s. of (5.4) implies that* $B_{a,b_1\cdots b_5} = 0$. $\tag{5.8}$

Some remarks are in order for these lemmas: First, Lemma 1 is based on the fact that each sector of different structure of $\gamma$-matrices in (5.1) for the two pair of indices $\alpha\beta$ and $\gamma\delta$ is to be independently zero. This leads to the condition $(3/8)A_{a,b} - (1/48)\eta_{ab}A_{c,}{}^c = 0$, whose trace gives $A_{c,}{}^c = 0$, which in turn yields $A_{a,b} = 0$, when re-substituted into this original equation. Second, Lemma 2 is also easy under (5.2), because we can require each of the two sectors with the $\gamma$-matrix structures $(\gamma^a)_{\alpha\beta}(\gamma^{bc})_{\gamma\delta}$ and $(\gamma^{[2]})_{\alpha\beta}(\gamma_{[2]}{}^{abc})_{\gamma\delta}$ should vanish independently. Note here that a simple contraction of two indices in (5.6a) does *not* lead to $B_{a,}{}^a{}_b = 0$, due to the vanishing trace component. We did not write the condition of vanishing of the second line on the r.h.s. of (5.2), because it is just a necessary condition of the conditions (5.6a) and (5.6b). We also mention that the well-known identity (2.11) is nothing else than a special case of $B_{a,bc} = \eta_{a[b}v_{c]}$ satisfying both (5.6a) and (5.6b). Third, Lemma 3 is straightforward, because each sector in the r.h.s. of (5.4) is to vanish independently. The last sector with $(\gamma^{[2]})_{\alpha\beta}(\gamma^{[2]'})_{\gamma\delta}$ yields

$$B_{a,}{}^a{}_{b_1\cdots b_4} = 0 \ , \tag{5.9}$$

which combined with the vanishing of the first and second lines of the r.h.s. of (5.4) implies immediately $B_{a,b_1\cdots b_5} = 0$. This is due to the difference in the coefficient in the first terms in these two sectors.

# 6. Concluding Remarks

In this Letter, we have shown that the $d=0$ $F$-BI can not be satisfied by any correction of the type $(\gamma^{ab})_{\alpha\beta}X_{ab}{}^c$ or $i(\gamma^{a_1\cdots a_5})_{\alpha\beta}X_{a_1\cdots a_5}{}^c$ in dimension zero supertorsion constraint $T_{\alpha\beta}{}^c$, based on the two assumptions: (i) M-theory corrections to $F$-superfield strength at $d<0$ are absent; (ii) The $T$ and $F$-BIs are not modified by Chern-Simons terms. Additionally, we relied upon the 'conventional constraints' that relate and delete unnecessary freedom in the expansion of $E_\alpha$ and $E_a$. These conventional constraints restrict the structure of corrections in $T_{\alpha\beta}{}^c$, such that the one-gamma term $i(\gamma^c)_{\alpha\beta}$ in $T_{\alpha\beta}{}^c$ receives no corrections, while the $\gamma^{[2]}$ or $\gamma^{[5]}$ terms can, satisfying the conditions in (2.7). We have seen that this result is valid for $F$-BI without Chern-Simons modifications.

Subsequently, we have also analyzed the BIs at $d \geq 1$, and obtained some conditions on the constraints of $T_{\alpha\beta}{}^\gamma$, $T_{ab}{}^c$ and $F_{\alpha bcd}$, at least for the case that the $F$-BIs are not modified by Chern-Simons terms. In particular, we have found that $F_{abcd} = 0$ in order to satisfy the matching exponential functions $e^\Phi$ in the $(a\beta\gamma,\delta)$-type BI at $d=3/2$. On the other hand, the consistency of gravitino superfield equation out of the same $(a\beta\gamma,\delta)$-type BI leads to the condition of vanishing of $\nabla_\alpha \Phi = 0$, leading to the trivial solution $\Phi = $ const. This validates our conclusion, because this scalar superfield $\Phi$ was the only possible modification at $d=0$. In other words, the modification of the constraint $F_{\alpha\beta cd}$ at $d=0$ is not enough for embedding M-theory corrections.

As has been mentioned, our result is not so surprising, but reasonable enough from the following viewpoints. Namely, the only Fierz identity of the form $(\gamma^{[m]})_{(\alpha\beta}(\gamma^{[n]})_{\gamma\delta)}$ is nothing other than (2.11). This is also understandable from the fact that the $F$-BI (2.3) at $d=0$ has four spinorial indices and one vectorial index as free indices, and therefore the vanishing of (2.3) gives such a strong condition as all the components in $U$'s and $X$'s are zero, except for the singlet $\mathcal{A}^{\{1\}}$ in $U_{abcd}$.

Our main conclusion in this paper can be bypassed by avoiding at least one of the two assumptions (i) and (ii) above. The assumption (i) seems very difficult to avoid, due to the fermionic symmetry of supermembrane action that seems to prevent the introduction of any $F$-constraints at $d<0$. On the other hand, the assumption (ii) is also difficult, because there has been no other known example of such supertorsion $T$-BIs modified by Chern-Simons terms. We mention also that our result here does *not* contradict the works in [8], because the $F$-BIs we dealt with in our paper has not been analyzed explicitly in [8]. In fact, a statement about the necessity of the $F$-constraints at $d<0$ was given in [8] without explicit proof. Our result in this paper provides a supporting proof with evidence for this statement, in terms of explicit computation of the $F$-BI at $d=0$.

We believe that not only the conclusion presented in this paper, but also the technical ingredient of Fierz identities will be of great importance in the future, for exploring any possible M-theory corrections into 11D superspace supergravity.

## Acknowledgements

We are grateful to M. Cederwall, S.J. Gates, Jr., U. Gran and B.E.W. Nilsson for considerable help and discussions for crucial points in this paper.